\documentclass[article,twocolumn,eqsecnum,showpacs]{revtex4}

\usepackage{dcolumn}
\usepackage{amsmath}
\usepackage{graphics}

\begin{document}

\title{Spontaneous decay of an atom excited in a dense and disordered atomic ensemble: quantum microscopic approach}
\author{A.S. Kuraptsev, I.M. Sokolov
\\
%EndAName
\small St.-Petersburg State Polytechnic University, 195251, St.-Petersburg, Russia
}

\date{\today}

\sloppy

%\affiliation

%\baselineskip22 pt\newpage

\begin{abstract}
On the basis of general theoretical results developed previously in [I. M. Sokolov et al., J. Exp. Theor. Phys. 112, 246 (2011)], we analyze spontaneous decay of a single atom inside cold atomic clouds under conditions when the averaged interatomic separation is less or comparable with the wavelength of quasi resonant radiation. Beyond the decay dynamics we analyze shifts of resonance as well as distortion of the spectral shape of the atomic transition.
\end{abstract}

\maketitle

\section{Introduction}
Influence of the environment on atomic spontaneous decay has attracted considerable interest both because of its fundamental importance and because of its significance in different applications such as quantum metrology, quantum information science, and lasing in disordered media \cite{7a}-\cite{7j}. These influences are of special interest for realization of highly stable and accurate optical atomic clocks \cite{7}-\cite{15}, particularly neutral-atom-based optical frequency standards \cite{15a}-\cite{15g}.

There are a great number of works where the environmental influence on the atomic decay constant is studied both theoretically and experimentally \cite{44a}-\cite{41}. In the vast majority of these researchs, the case of a transparent and continuous surrounding medium is considered.  For theoretical description of such systems, both macroscopic and microscopic approaches are used. Depending on the different assumptions about the local environment of the embedded atom several models have been proposed and studied.  Among the most important, we note such approaches as a virtual cavity, a real cavity and so-called "fully microscopic" models.

The main goal of the present work is to study theoretically the case of disordered atomic ensembles with averaged interatomic distances greater than or comparable with the wavelength of the atomic transition as it is held in the optical atomic clock based on cold atoms \cite{15a}-\cite{15g}.  In such a case discrete structure of atomic clouds should be taken into account. Besides that the atomic radiation is quasi-resonant, and a model assuming quasi-transparent non-absorbing media is not valid.

In this work we make numerical Monte-Carlo simulations based on a quantum microscopic approach \cite{44}. Numerical simulation allows us to analyze the process of single atom excitation decay, taking into account resonant multiple incoherent scattering of the light inside atomic cloud, without introduction of any adjustable parameters, as opposed to approaches used earlier for similar problems in  \cite{38}-\cite{41}. In addition, we do not use a continuous media approximation. It allows us to take into account correctly the discrete structure of atomic clouds and consequently the interatomic correlations which play, in our case, an important role.

One further unique feature of the present work is that we calculate not only the decay constant, but also dynamics of decay and the spectrum of the atomic transition, particularly the shifts of atomic lines caused by the dipole-dipole interaction. The question of density dependent spectral line shifts for cold atomic gases recently gave rise to a wide discussion, \cite{45a}-\cite{45f} and this question is especially important for optical frequency standards \cite{15d}, \cite{45r}.

In the following paragraphs, we first lay out our basic assumptions and approach. This is followed by a presentation of our main results, and a discussion of them.

\section{Basic assumptions and approach}

The calculation of spontaneous decay dynamics in this paper will be made on the basis of a microscopic quantum approach developed in \cite{44}. A similar approach was previously used for analogous problems in \cite{27}, \cite{28}. The authors of these papers searched for analytical solutions, in order to have an opportunity to analyze the validity of real and virtual cavity approximations.  They succeeded in finding an approximate expression for the decay constant up to the second order of the parameter $n\alpha$, where $n$ is the atomic density and $\alpha$ is the polarizability of a free atom. Besides that, final expressions in these papers were obtained for transparent and continuous medium.

In the present work we search for numerical solutions to this problem. It allows us to obtain results in all order of $n\alpha$ and for arbitrary parameters of the excited atom.  The approach used is depicted in detail in \cite{44} and we will not discuss it here.  We will mention only the main approximations used hereafter and will show the principal analytical expressions utilized for numerical calculations.

Note also, that a similar microscopic approach combined with numerical calculation has been used in \cite{PC10} for determination of averaged decay rate and for discussion the influence of local density of collective states on spontaneous decay of impurity atoms. However the authors did not analyze the explicit time dependence of atomic population and did not consider the spectrum of the atomic transition.

We consider an ensemble consisting of N atoms. All atoms have a ground state with $J = 0$ and an excited level with $J = 1$.  In the present paper we analyze two different cases. In the first, all atoms including the initially exited one are identical. The physical properties of clouds considered in this part of the paper correspond to the situation realized in atomic clocks \cite{15a}-\cite{15g}. Namely, interatomic separations are relatively big and the discrete structure of the clouds is important. The identity of the atoms means that radiation emitted by one of them is quasi resonant for the other. The main goal of this part of the paper is to study the single atom decay under such conditions.
By now single atom spectral response has not been studied in experiment. In typical experiments external radiation interacts with whole (or essential part of) atomic clouds. However in our opinion such a problem statement is interesting from a theoretical point of view. Beside that it is not completely abstract. Local excitation of atoms inside optically depth and dense clouds can be performed by means of method suggested earlier by M. D. Havey \cite{PC}.

For such excitation one has to illuminate the cloud with two narrow and off resonant orthogonally propagated light beams. Each beam does not cause single photon excitation; but their simultaneous interaction with atoms in the crossing region can cause two-photon excitation from the ground $S$ to excited $D$ state if conditions of two-photon resonance are satisfied.  The subsequent spontaneous transition from $D$ to $P$ state leads to population of the studied $P$ state. Thereby described method allows obtaining small cluster of excited atoms in the middle of the cloud. For simplicity thereafter in the paper we will consider the case of only one atom excited. Note that possibilities of two-photon excitation $5s\; S-2(1/2)\rightarrow 5p\; P-2(j)\rightarrow 5d\; D-2(j)'$ of rubidium atoms have been already studied in experiment \cite{JPB06}.

The second case we will consider in the paper corresponds to spontaneous decay of an impurity atom embedded in homogeneous ensemble. In this case the transition frequency of the embedded atom $\omega_{emb}$  and the natural linewidth of its excited state $\gamma_{emb}$  differ from the corresponding values  $\omega_{0}$ and $\gamma_{0}$   of surrounding atoms.

In our calculation we will assume that all atoms are motionless. To take into account residual atomic motion and the random spatial inhomogeneity of the atomic ensemble, we will consider the statistical ensemble of clouds with a random distribution of atoms. The results presented below are obtained by averaging over this ensemble using the Monte Carlo method.

The microscopic approach enables us to consider the clouds of arbitrary form with an arbitrary nonuniform spatial distribution of atoms.  The initially excited atom can also be located at an arbitrary position.  Further, we will assume that atoms surrounding the emitting atom are randomly distributed in a cubic volume. The random distribution is uniform.  The initially excited atom itselfis in the center of the cube.  For the chosen geometry the system is isotopic on average, so we can consider the spontaneous decay of any Zeeman sublevel.  For determinacy we will assume that at initial time $t=0$ only one substate $m=-1$ of the central atom is populated. All the other atoms of the ensemble are in their ground state having $J=0,\;m=0$.

All approximations described above allow us to avail ourselves of the results of the general theory developed in \cite{44}. This theory was then mainly used for description of the interaction between light and dense atomic clouds in steady-state conditions \cite{45},  \cite{46}-\cite{48}. Here we analyze the dynamics of the dense atomic ensemble.

According to \cite{44}, the time dependent amplitudes of the collective atomic states with one excited atom $b_{e_a^m}(t)$ can be found as follows

\begin{equation}
b_{e_a^m}(t)= \int\limits_{-\infty }^{\infty }\frac{d\omega
}{2\pi}exp(-i\omega t)\sum\limits_{e_b^{m'}}
R_{e_a^m e_b^{m'}}(\omega)b^0_{e_b^{m'}}. \label{2}
\end{equation}%
Here, the index $e_a^m$ contains information about the number $a$ of atom excited in any considered state. It also indicates the specific Zeeman sublevel $m$ which is populated.

The vector $b^0_{e_b^{m'}}$ is determined by the initial excitation of the entire ensemble. In the considered case this vector contains only one non-zero element corresponding to the Zeeman sublevel $m'=-1$ of the central atom.

The matrix $R_{e_a^m e_b^{m'}}(\omega)$ in Eq. (\ref{2})  takes into account the possible excitation of the atoms by the reradiated light. As it was shown in \cite{SKKH09}, $R_{e_a^m e_b^{m'}}(\omega)$ is a resolvent operator of the considered system projected on the states consisting of single atom excitation, distributed over the ensemble, and the vacuum state for all the field modes
\begin{equation}
R_{e_a^m e_b^{m'}}(\omega )=\left[ (\omega -\omega _{a})\delta _{e_a^m e_b^{m'}}-\Sigma
_{e_a^m e_b^{m'}}\right] ^{-1},  \label{4}
\end{equation}
Here $\omega _{a}$ is equal to $\omega _{emb}$ for the embedded atom and $\omega _{0}$ for surrounding atoms.

If $a\neq b$, the matrix $\Sigma_{e_a^m e_b^{m'}}$ describes the excitation exchange between atoms $a$ and $b$ inside the cloud. In this case its matrix elements are
\begin{eqnarray}
&&\Sigma _{e_a^m e_b^{m'}}=\sum\limits_{\mu ,\nu}
\frac{\mathbf{d}_{e_a^{m};g_{a}}^{\mu }\mathbf{d}_{g_{b};e_b^{m'}}^{\nu }}{\hbar r^{3}}\times \label{5} \\&& \left[ \delta _{\mu \nu }\left(
1-i\frac{\omega _{a}r}{c}-\left( \frac{\omega _{a}r}{c}\right) ^{2}\right)
\exp \left( i\frac{\omega _{a}r}{c}\right) \right. -
\notag  \\&&
\left. -\dfrac{\mathbf{r}_{\mu }\mathbf{r}_{\nu }}{r^{2}}\left( 3-3i\frac{\omega _{a}r}{c}-\left( \frac{\omega _{a}r}{c}\right) ^{2}\right) \exp
\left( i\frac{\omega _{a}r}{c}\right) \right] .
\notag
\end{eqnarray}

Here $\mathbf{r}_\mu$ is the projection of the vector $\mathbf{r}=\mathbf{r}_{a}-\mathbf{r}_{b}$ on the axis $\mu$ of the chosen coordinate frame and $r=|\mathbf{r}|$ is the interatomic distance. Note that the expression (\ref{5}) is written in the so-called pole approximation. This approximation is based on the fact that (\ref{5}) depends very slowly on $\omega_a$, so in calculating (\ref{5}) we can consider that for all $a$ $\omega_a=\omega_0$.

If $a$ and $b$ are the same atom then $\Sigma _{e_a^m e_b^{m'}}$ differs from zero only for $m=m^\prime$. In this case $\Sigma _{e_a^m e_a^{m}}$ determines the Lamb shift and the decay constant of the corresponding excited state. Including Lamb shifts in the transition frequency $\omega _{a}$ we get
\begin{equation} \Sigma _{e_a^m e_a^{m'}}=-i\delta_{mm'}\gamma_a /2.
\label{6}
\end{equation}

In the next section, we use relations (\ref{2})-(\ref{6}) to calculate both the time dependence of the excited state population and the spectrum of the atomic transition.

\section{Results and discussion}
\subsection{Spontaneous decay dynamics}
Relations (\ref{2})-(\ref{6}) allow calculating the time-dependent population of any Zeeman sublevel of any atom in an ensemble. Here we will analyze the decay of an initially populated sublevel $J=1,\;m=-1$ of central atom. For brevity we denote the amplitude of this state as $b_{0}(t)$. The corresponding population $P_0(t)$ can be found as follows

\begin{equation} P_{0}(t)=<|b_{0}(t)|^{2}>. \label{3}
\end{equation}

Here angle brackets denote averaging over a uniform random atomic distribution of surrounding atoms. As was mentioned previously, this averaging is performed by a Monte Carlo method.

Besides $P_0(t)$ we will calculate the time dependent decay rate $\gamma(t)$

\begin{equation} \gamma(t)=-\frac{1}{P_{0}(t)}\frac{dP_0(t)}{dt}. \label{4}
\end{equation}

Figure 1 shows the spontaneous decay dynamics of an excited atom in the case when the initially excited atom coincides with all other atoms in the cloud, i.e. in the case when $\gamma_{emb}=\gamma_{0}$ and $\Delta=\omega_{emb}-\omega_0=0$. The calculations were made for an atomic density $n=0.05$ and for three different $N$ -- $N=300$; $N=1000$ and $N=2000$. Hereafter in this paper we use the inverse wave number of the resonance probe radiation in vacuum $k_0^{-1}=c/\omega_{0}$ as a unit of length.

\begin{figure}
\begin{center}
{$\scalebox{0.4}{\includegraphics*{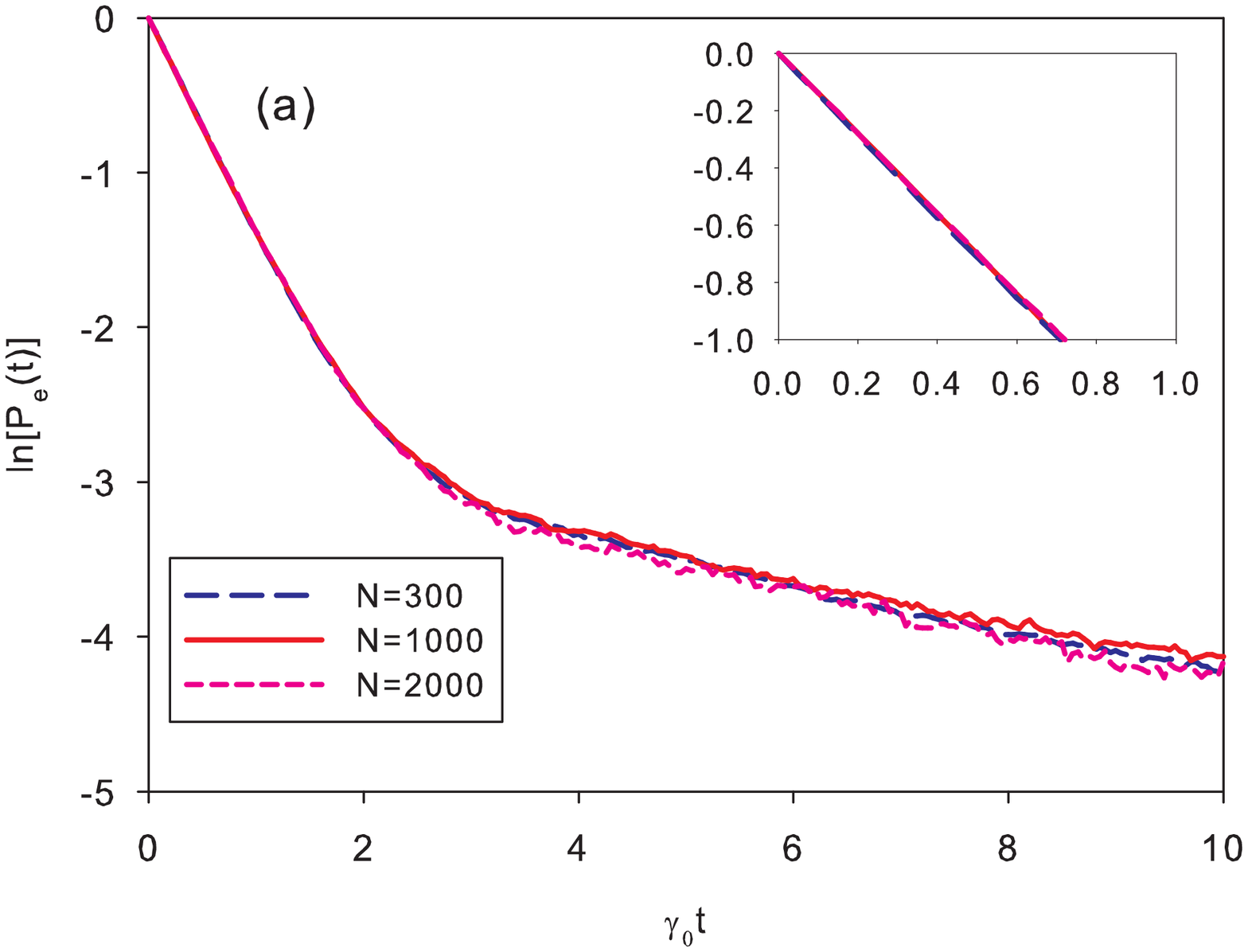}}$ }{$\scalebox{0.4}{%
\includegraphics*{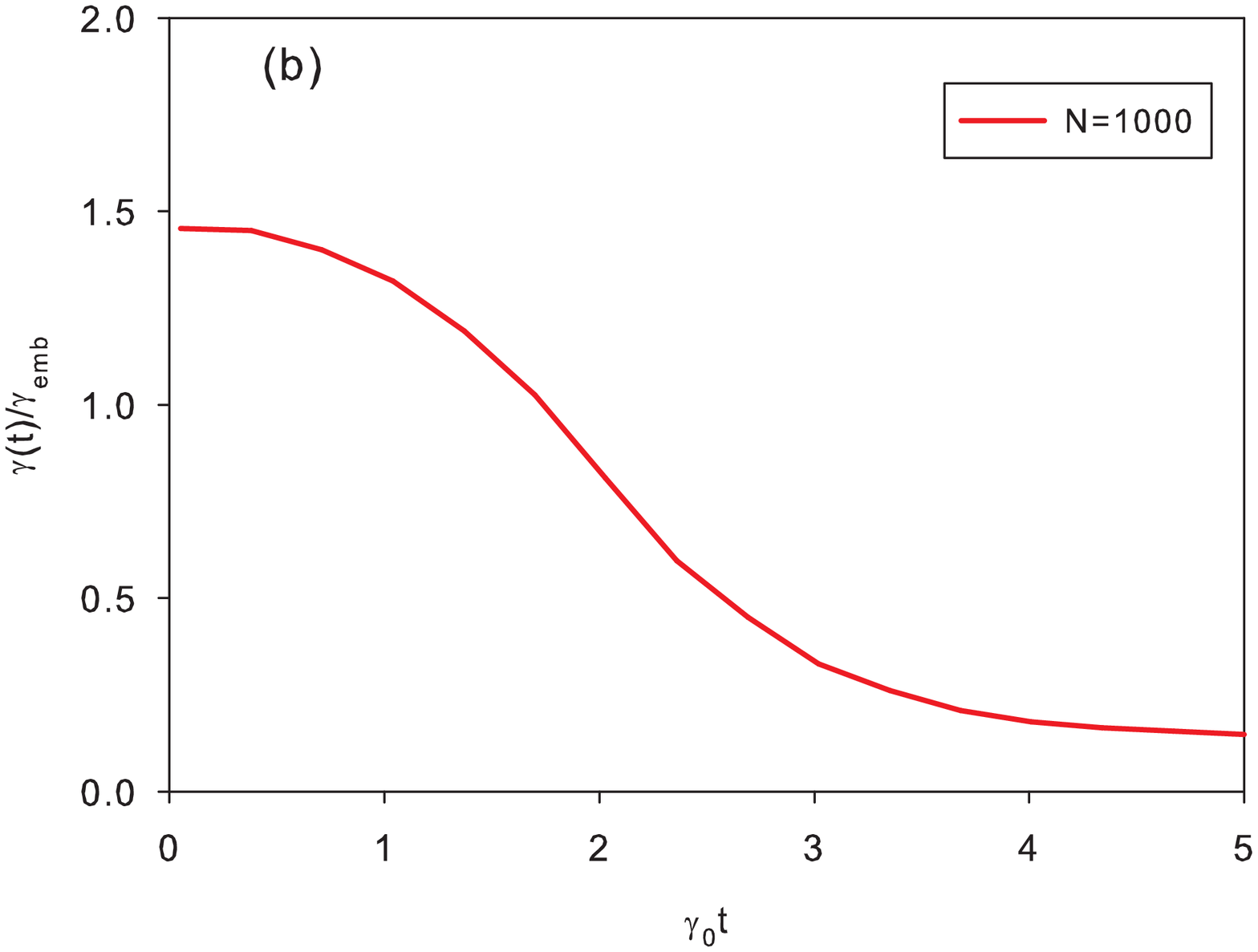}}$ }
\caption{(Color online) Spontaneous decay dynamics (a); time dependence of the effective spontaneous decay rate (b); $n=0.05$, $\gamma_{emb}=\gamma_{0}$, $\Delta=0$}
\end{center}
\par
\label{fig1}
\end{figure}

Figure 1a demonstrates that the process of spontaneous decay in a dense atomic ensemble is described by a multi-exponential law. Initial state with only one excited atom can be expanded as a series over set of nonorthogonal eigenvectors of Green matrix for each random spatial configuration of the system. Among different collective quantum states in the considered ensemble there are both super- and sub-radiant ones. The former influences the spontaneous decay dynamics predominantly in the early stages; an atom inside the ensemble decays faster than a free atom. Even in the case of relatively small densities like $n=0.05$ the increasing in decay rate is important.  With time the role of sub-radiant states increases, resulting in a decrease in decay rate. For $t>1/\gamma_{emb}$ this rate becomes less than that for free atoms (see Fig. 1b).

The specific type of decay dynamics for given density depends on the size of the atomic ensemble, i.e. on the number of atoms N in it. It is typical situation when sub and super radiant states manifest themselves in collective decay (for more detail see \cite{AGK08}-\cite{BPK12}). We have analyzed how the function $P_0(t)$ changes with $N$ and found that for small clouds when mean free path of photon less or comparable with linear size of atomic ensemble these changes are very essential. As $N$ and linear size increase the changes in $P_0(t)$  become more and more weak. This dependence has evident tendency to saturation.  Our calculation shows that for studied geometry for $n=0.05$ the curve $P_0(t)$ does not practically changes as $N$ becomes more than 300. So results obtained for $N>300$ can be used for description of single atom spontaneous decay inside any macroscopic ensemble with reasonable accuracy.

It is illustrated by the fig. 1a where we show time dependence of $P_0(t)$ for three different $N$. For times less or comparable with $3/\gamma$ all three curves coincide with accuracy of calculations. Such behavior connects with specific initial condition considered in this paper (contrary to the cases studied in \cite{AGK08}-\cite{ BPK12}). Modification of decay rate of initially excited central atom is caused by radiation of this atom which is scattered back by atoms of environment. Results displayed in the fig.1a show that atoms of the cloud located far from central atom influence on decay faintly at this time interval. Slight discrepancy at bigger times when population $P_0(t)$ is already small, connects with the fact that lifetimes of the most long-lived collective states are very sensitive to specific spatial configuration of the cloud and for accurate averaging of corresponding dates volume of calculation  has to be increased immensely.

Increasing the density of surrounding atoms causes enhancement of collective effects and consequently more considerable modification of spontaneous decay.  This effect is illustrated by Fig. 2 which shows the dynamics of single atom excitation decay inside an ensemble with the atomic density $n=1$.

\begin{figure}
\begin{center}
{$\scalebox{0.4}{\includegraphics*{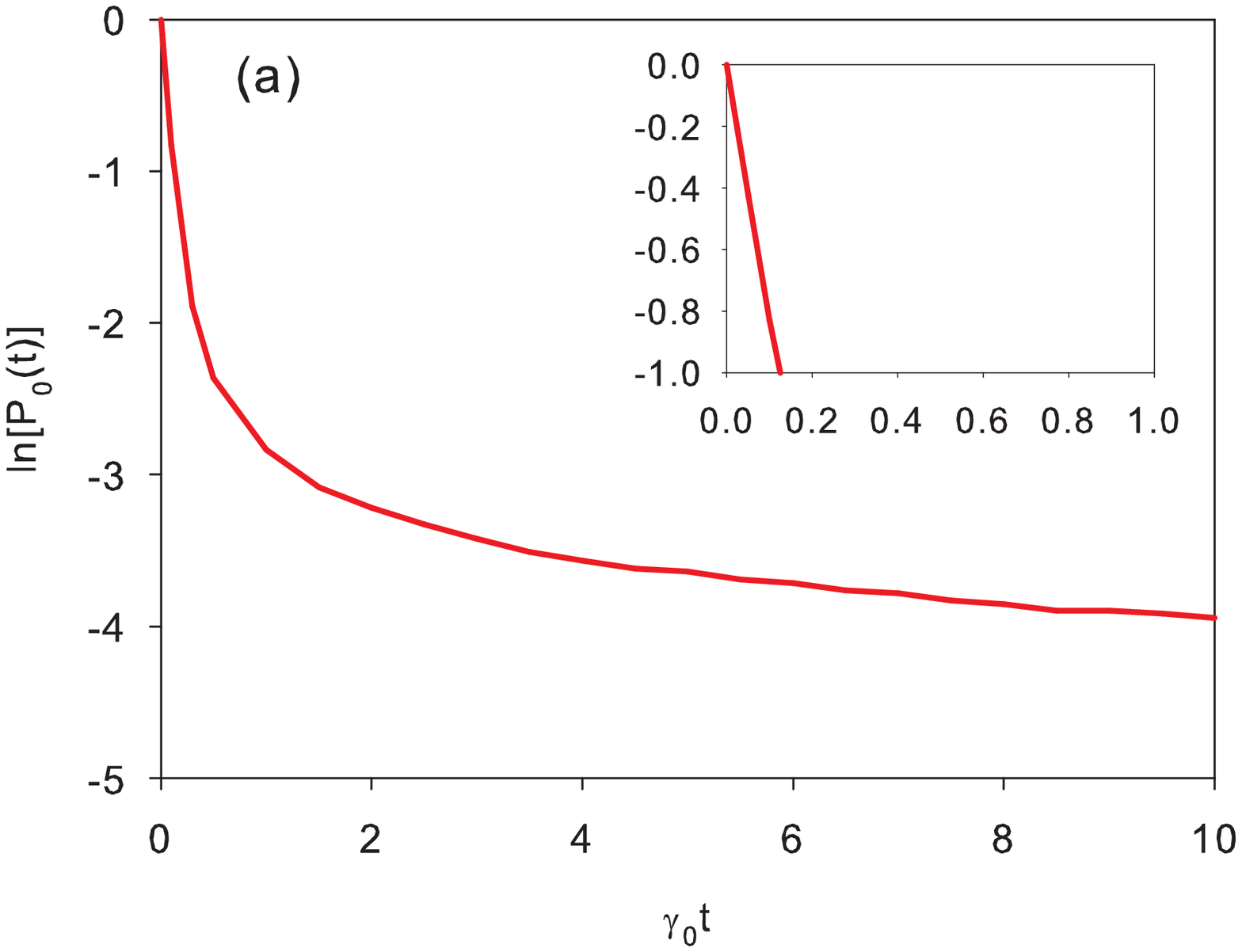}}$ }{$\scalebox{0.4}{%
\includegraphics*{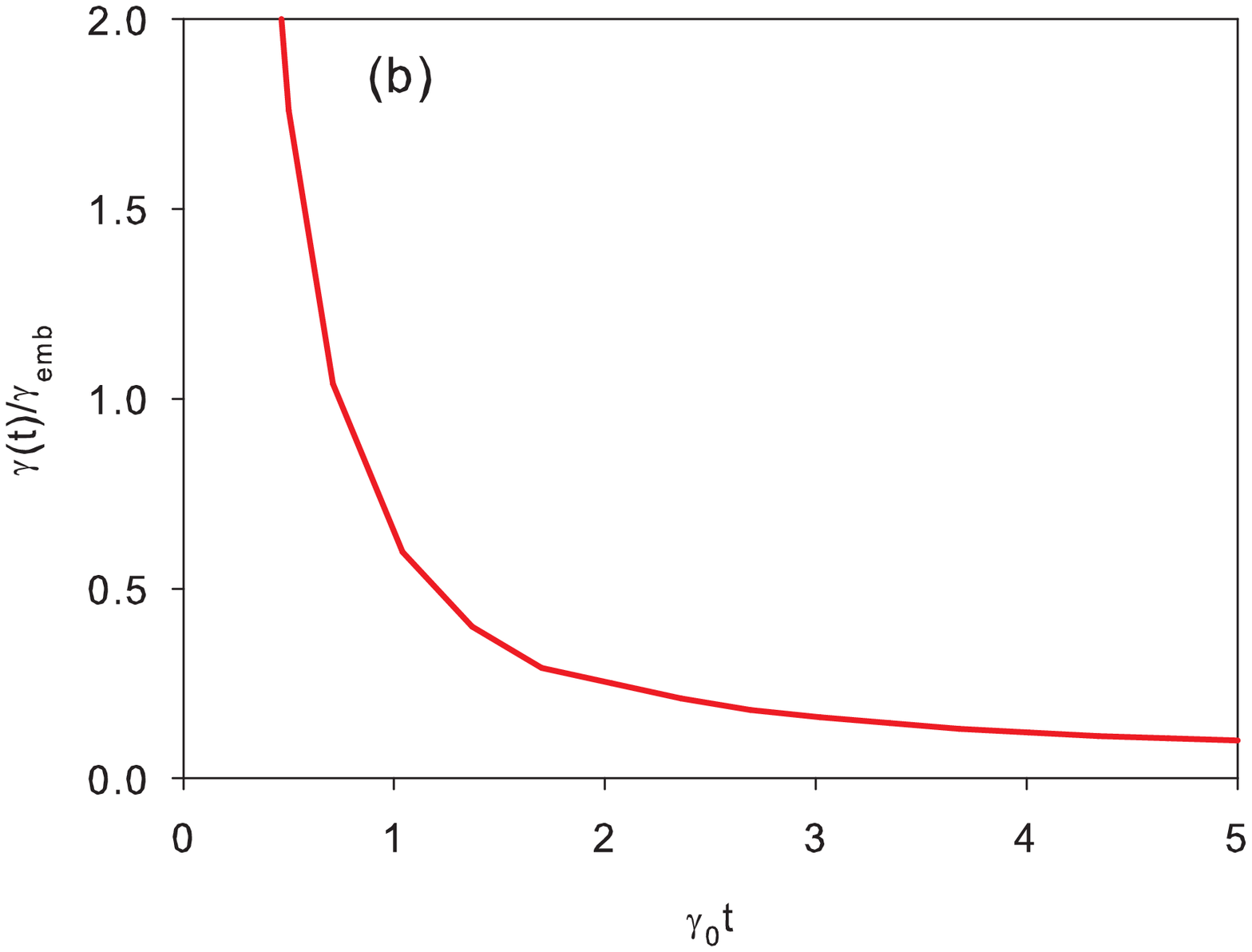}}$ }
\caption{(Color online) Spontaneous decay dynamics (a); time dependence of the effective spontaneous decay rate (b); $n=1$, $\gamma_{emb}=\gamma_{0}$, $\Delta=0$, $N=2000$}
\end{center}
\par
\label{fig2}
\end{figure}

At short times the decay constant is several times greater than for free atoms.  On the contrary, for $t>1/\gamma_{emb}$ the decay process shown in Fig. 2 goes noticeably slower than in case of $n=0.05$.

As density increases the size of mesoscopic cloud for which we can neglect dependence $P_0(t)$ on $N$ is also increase. For $n=1$ we observed saturation for $N\approx 2000$.

Figures 1 and 2 describe one atom excitation decay in completely homogeneous clouds.  In this case, the frequency of emitted photon lies in the absorption band of the surrounding atoms. Consider now the case of spontaneous emission of a foreign atom embedded in a dense ensemble when this atom and atoms of the medium have essentially different transition frequencies.

We will choose parameters of the cloud's atoms in such a way that the dielectric constant would be real with a good accuracy.  It gives us possibility to compare our results obtained in the framework of our quantum microscopic approach with different models suggested earlier.

We can neglect imaginary part of dielectric constant if $Im(\varepsilon(\omega))\ll Re(\varepsilon(\omega))-1$ for all frequencies emitted by the impurity atoms. It is known that for large detuning from resonance $Re(\varepsilon(\omega))-1$ is inversely proportional to $\Delta$ whereas $Im(\varepsilon(\omega))$ is inversely proportional to $\Delta^{2}$. So to have the possibility to satisfy corresponding inequality we have to choose $\omega_{emb}-\omega_0$ several times greater than $\gamma_{0}$. The width of emitted radiation should be relatively small so that necessary inequality would be satisfied not
only at $\omega_{emb}$ but at all the spectral area of the radiation.

The detailed analysis performed on the basis of known dielectric permittivity of cold atomic ensembles \cite{45}, \cite{48} shows that the requirements formulated above are satisfied particularly for  $\Delta=4\gamma_{0}$ and $\gamma_{emb}=0.1\gamma_{0}$ for the atomic density $n=0.05$; and for $\Delta=70\gamma_{0}$ and $\gamma_{emb}=2\gamma_{0}$ for the atomic density $n=1$.

Figure 3 shows the spontaneous decay dynamics of an atom embedded in an ensemble of atomic density $n=0.05$ and $n=1$ obtained in the framework of the quantum microscopic approach.  For comparison we included results predicted for the same parameters by a real cavity model ($\gamma_{rc}=\sqrt{\varepsilon}\left( 3\varepsilon/(2\varepsilon+1)\right) ^{2}\gamma_{emb}$), a virtual cavity ($\gamma_{vc}=\sqrt{\varepsilon}\left( (\varepsilon+2)/3\right) ^{2}\gamma_{emb} $) and a so-called "fully microscopic" model ($\gamma_{fm}=\left( (\varepsilon+2)/3\right)\gamma_{emb}$)

\begin{figure}
\begin{center}
{$\scalebox{0.38}{\includegraphics*{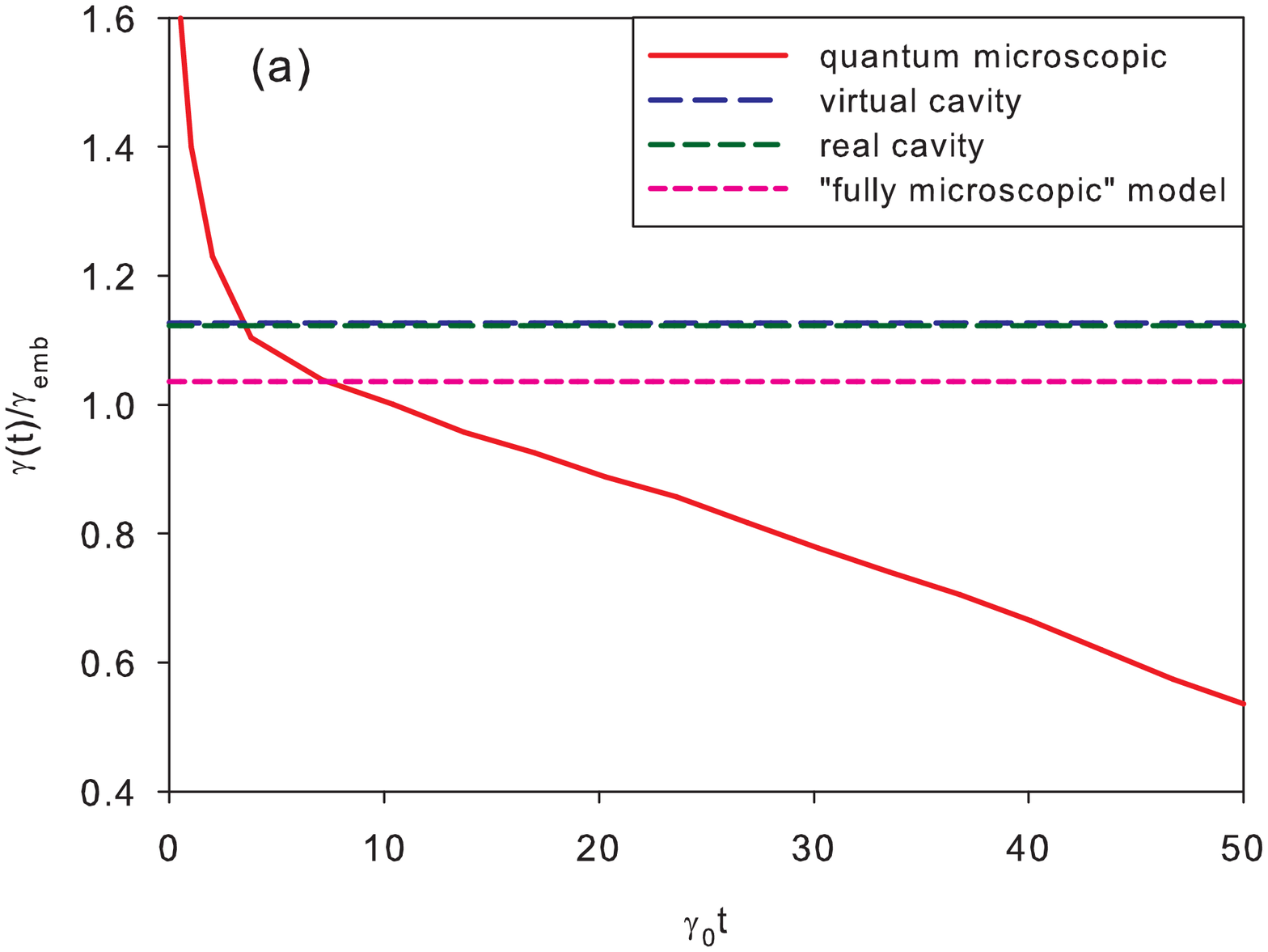}}$ }{$\scalebox{0.38}{%
\includegraphics*{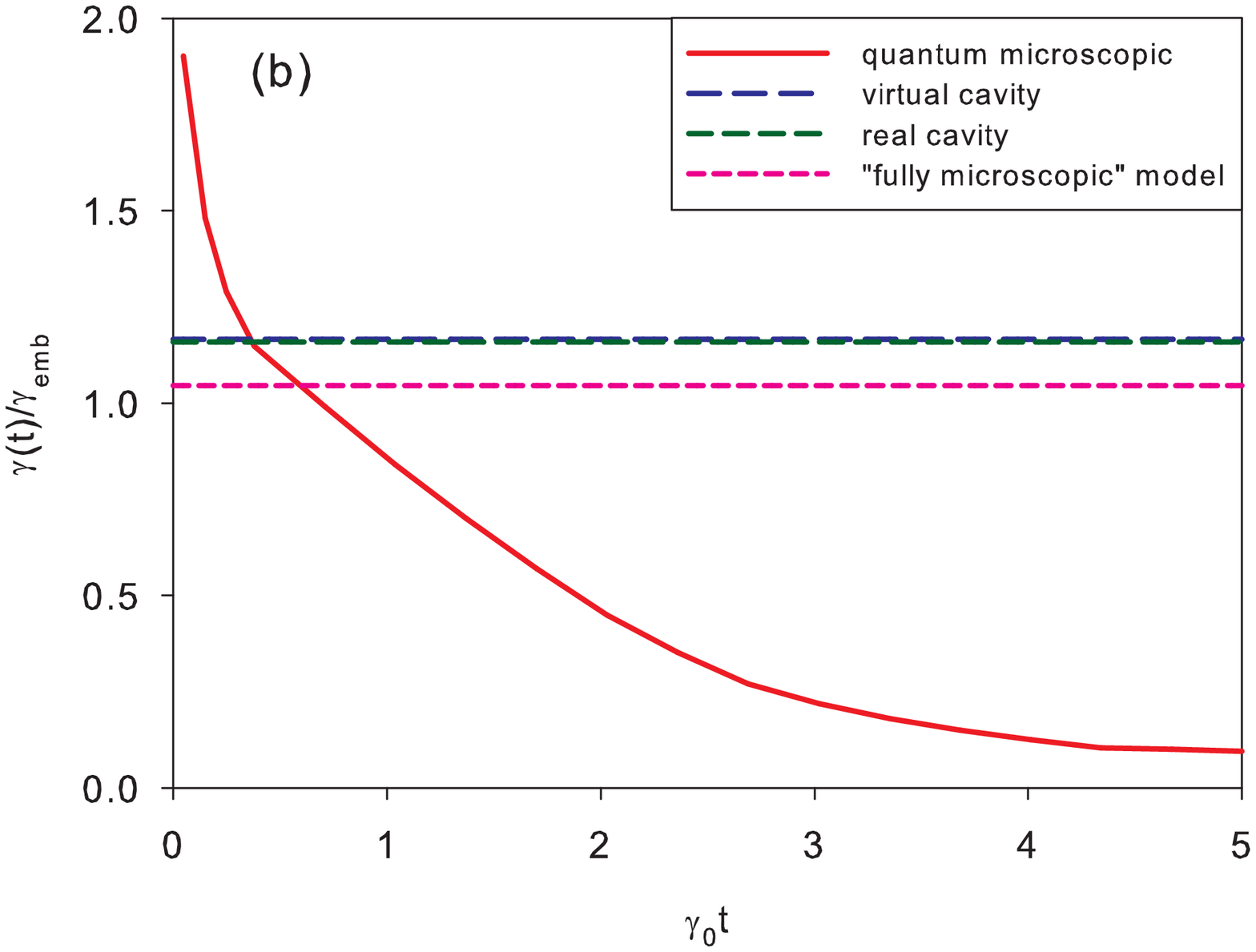}}$ }
\caption{(Color online) Time dependence of the effective spontaneous decay rate (a): $n=0.05$, $\gamma_{emb}=0.1\gamma_{0}$, $\Delta=4\gamma_{0}$; (b): $n=1$, $\gamma_{emb}=2\gamma_{0}$, $\Delta=70\gamma_{0}$}
\end{center}
\par
\label{fig3}
\end{figure}

It is clear for both considered cases that spontaneous decay of the foreign atom is described by a multi-exponential law. The current value of the decay constant depends on time.  There is an essential difference between predictions of our quantum approach and the different models suggested earlier. In our opinion the main reason for such a discrepancy lies in using a continuous medium approach. This approach does not allow one to take into account inter-atomic correlations quite correctly.

\subsection{Atomic transition spectrum}
Interaction of a radiating atom with its surroundings causes not only modification of the decay constant but also a shift of the atomic transition. Moreover, this interaction can be the reason of fundamental distortion of the spectral line.
At present there is a whole number of theoretical works devoted to analysis of spectral properties of cold atomic clouds. Theory predicts that different observables have different spectral dependence for dense clouds and these dependencies transform differently as density increases. Thus, the spectrum of the dielectric permittivity of cold and dense atomic clouds has a blue shift \cite{45}, \cite{45h}.  The maximal probability to excite the dense ensemble is observed for a negative detuning of exciting radiation \cite{46}.  The calculation of the total scattering cross section shows that the corresponding spectrum has several resonances \cite{44}.  Two of them are red shifted, and the third is in the blue region. Spectral dependence of fluorescence also has complicated behavior. It depends both on direction of fluorescence and its polarization \cite{44}.
By now density dependence of spectral properties of cold clouds has been studied in several experiments \cite{15d}, \cite{43},\cite{43a}.  In these experiments laser induced fluorescence of atomic ensemble was studied. Experiments showed red shift and some distortion of excitation spectrum.

In experiment external laser radiation excites whole atomic ensemble. Different atoms in the cloud excite with different probabilities. Observed fluorescence is result of averaging of secondary radiation of all atoms. And it depends essentially on specific type of excitation.  For typical spatially inhomogeneous clouds main contribution comes from relatively dilute external regions. In such a case observed collective effects are weak and spatial inhomogeneity makes interpretation of experimental result more difficult.

In this paper we analyze decay of local excitation inside dense atomic ensemble. The transition spectrum can be described by inverse Fourier transformation of $b_{0}(t)$ (\ref{2}).  It is completely determined by Fourier components of the resolvent matrix (\ref{4}).

\begin{figure}
\begin{center}
{$\scalebox{0.38}{\includegraphics*{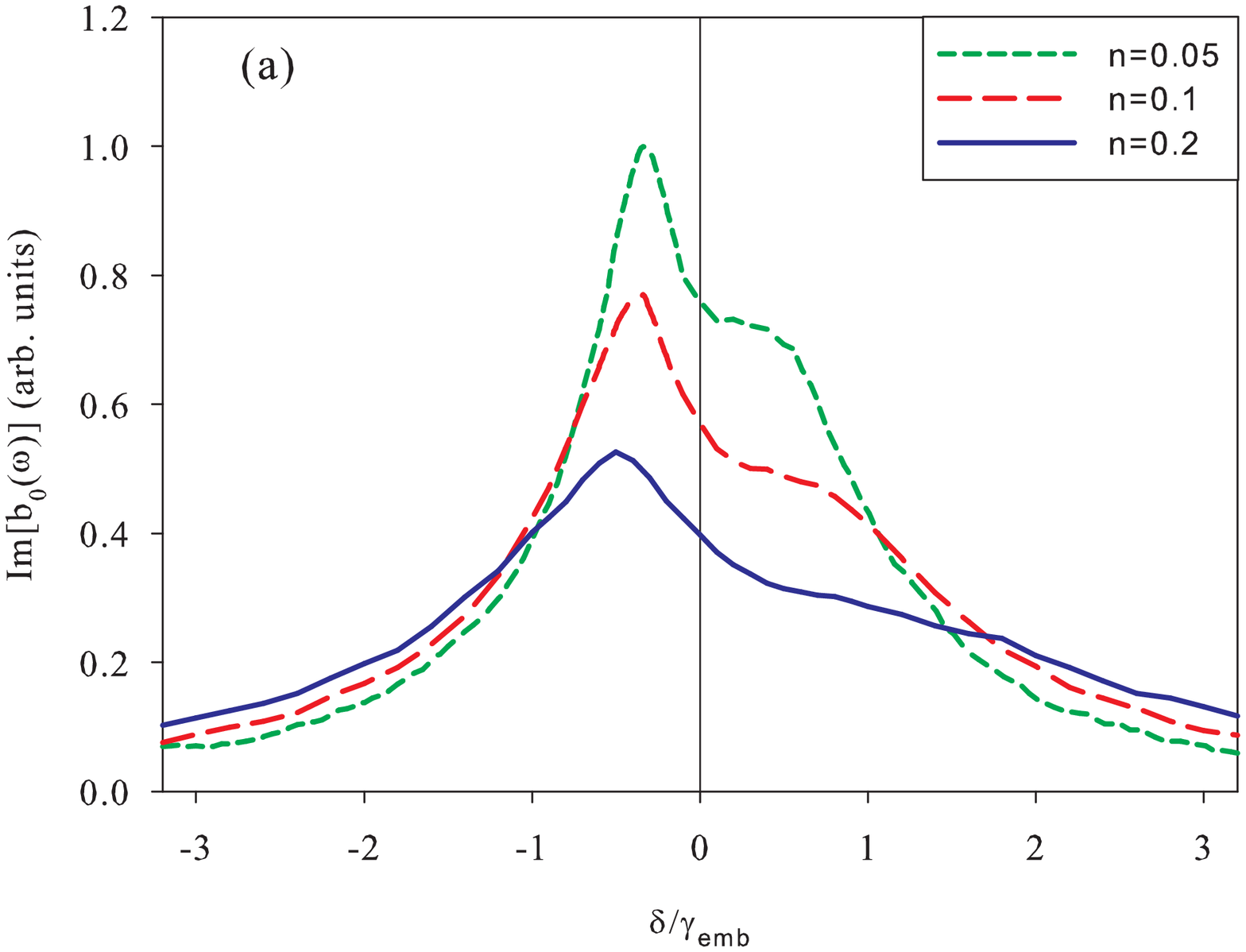}}$ }{$\scalebox{0.38}{%
\includegraphics*{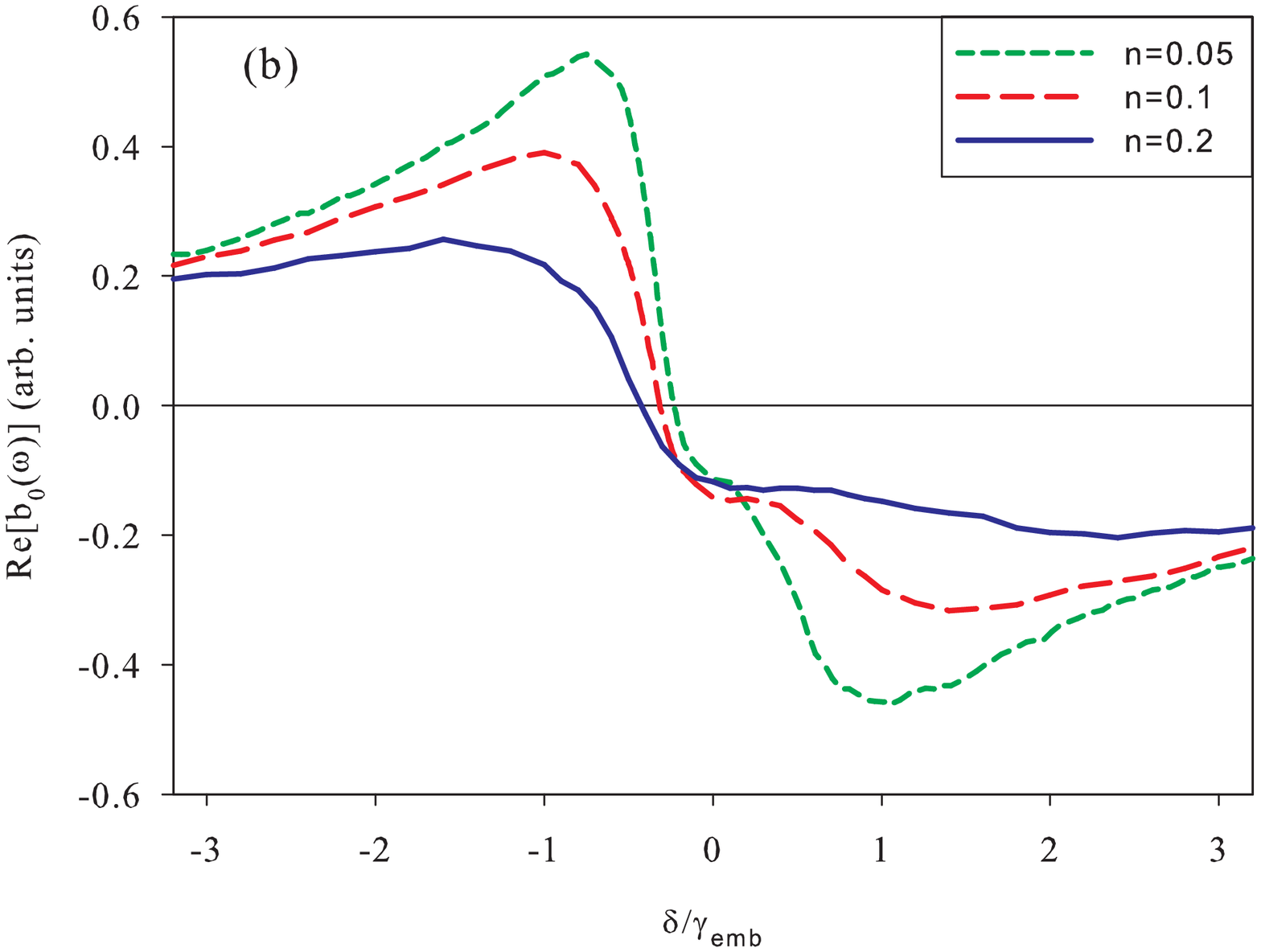}}$ }
\caption{(Color online) Atomic transition spectrum for different densities of atomic ensemble: absorbtion curve (a); dispersion curve (b)}
\end{center}
\par
\label{fig4}
\end{figure}

Figure 4 shows the spectrum of an atomic transition in the case when the excited atom is identical to the surrounding ones.
It demonstrates that the absorption curve (Fig. 4.a) as well as the dispersion one (Fig. 4.b) are significantly modified in comparison with the classical Lorenz and dispersion curves typical for a free atom. The contours have relatively narrow peaks and wide wings. The maximum of the spectral line is shifted in lower frequency as compared with a free atom. The value of the shift and average width of the contour increases with density.

In our opinion one atom spectral response analyzed above can be the foundation of new practical method of studying of collective effects in dense media. Few atom excitations can be realized in experiment as it described in sec.II. It gives opportunity to create local excitation in deep part of dense ensemble and study directly single atom polarizability modified by resonant dipole-dipole interaction.  It will give additional information for further analysis of complicate influence of collective effects on the spectral properties of cold clouds. Relatively strong modification of spectrum allows us to hope that this method will be sensitive.

In Fig. 5 we show the transition line shapes of a foreign atom embedded in a homogeneous atomic ensemble. Distortion of the line shape here is noticeably weaker than in the case shown in Fig. 4. Such an effect is directly connected with weakening of dipole-dipole interaction between the embedded atom and atoms of the surrounding ensemble as the difference in resonance frequencies of these atoms increases.

\begin{figure}
\begin{center}
{$\scalebox{0.38}{\includegraphics*{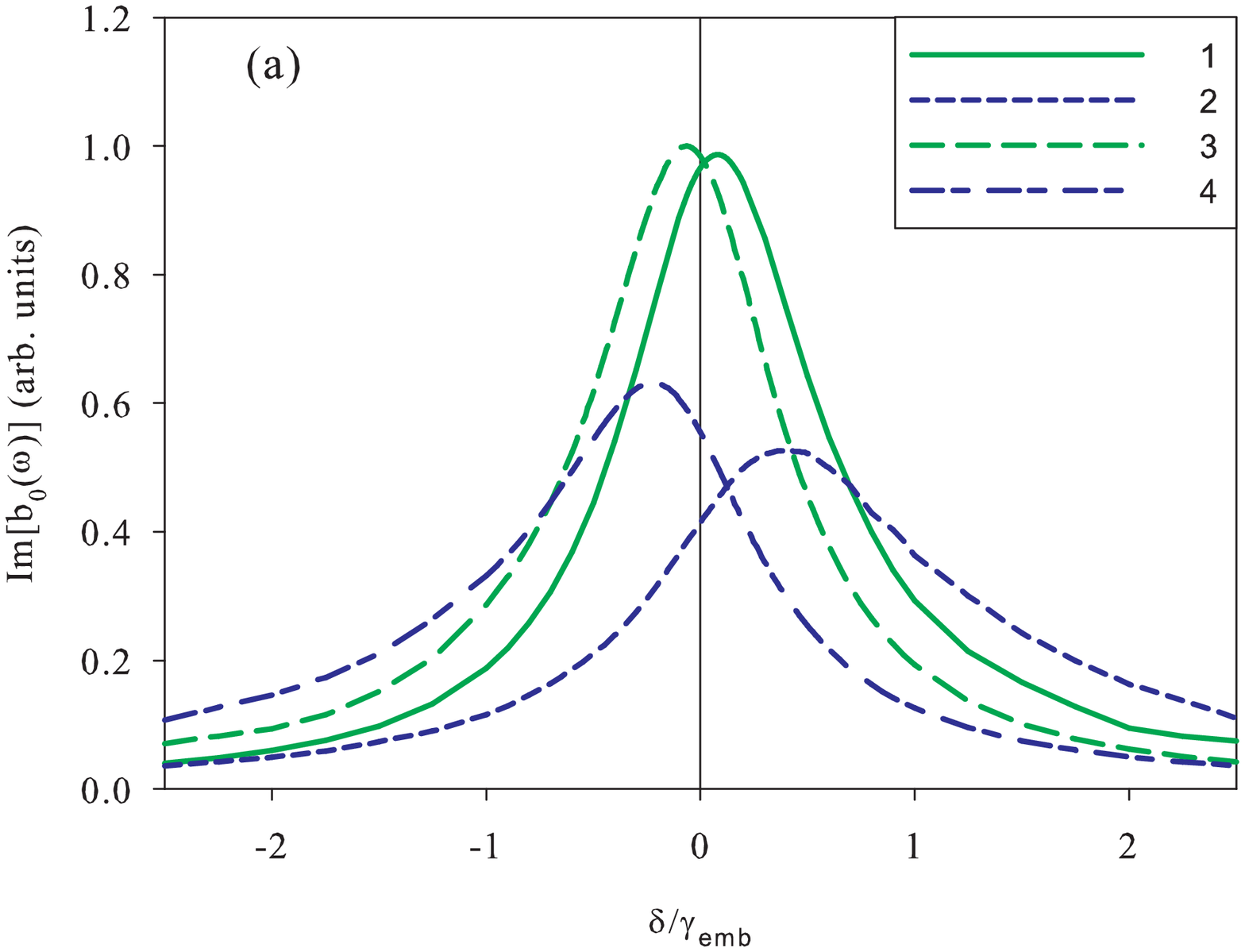}}$ }{$\scalebox{0.38}{%
\includegraphics*{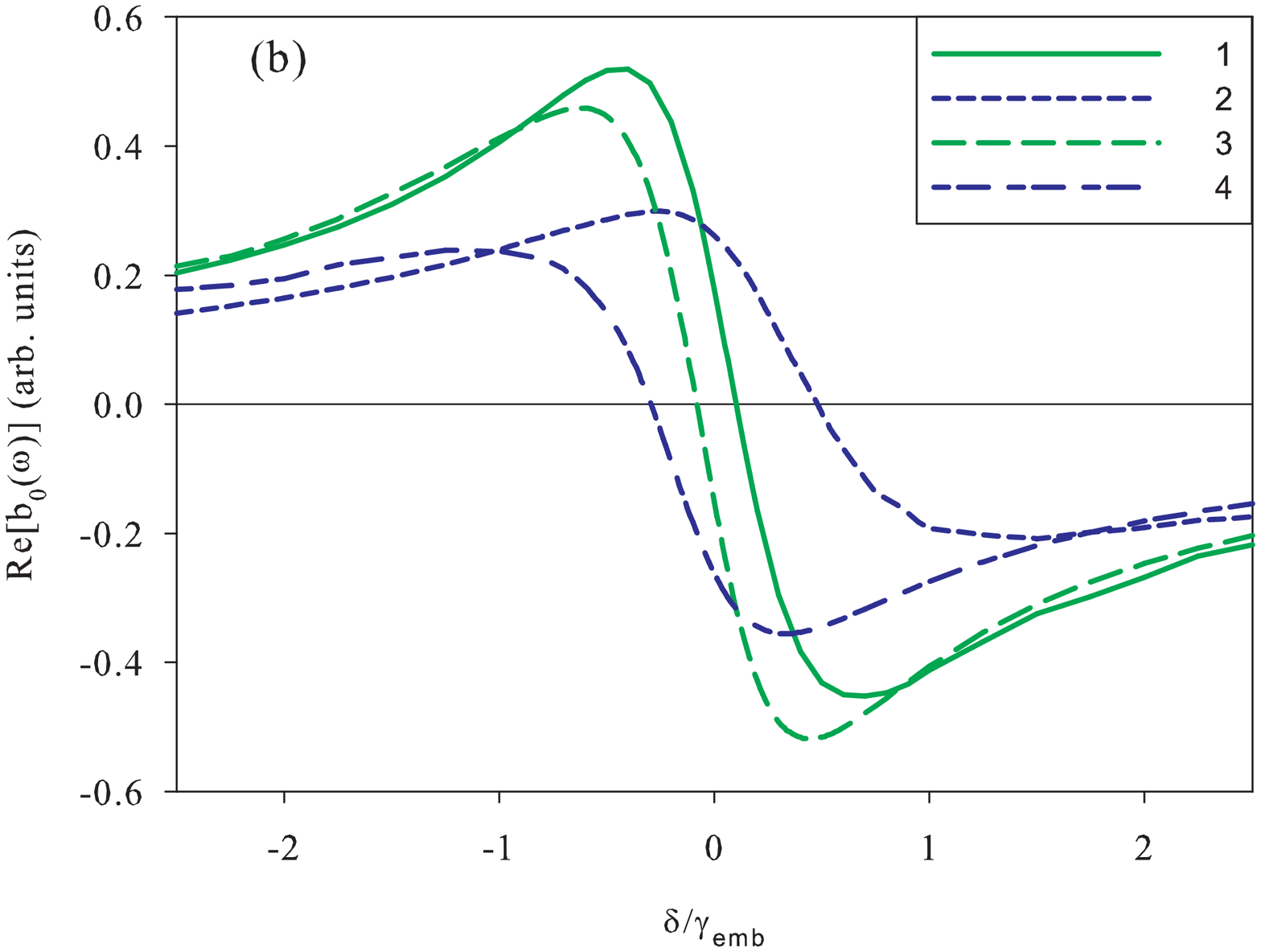}}$ }
\caption{(Color online) Atomic transition spectrum: absorbtion curve (a); dispersion curve (b); $\gamma_{emb}=0.1\gamma_{0}$; 1 -- $n=0.05,\;\Delta=+4\gamma_0$; 2 -- $n=0.2,\;\Delta=+4\gamma_0$; 3 -- $n=0.05,\;\Delta=-4\gamma_0$; 4 -- $n=0.2,\;\Delta=-4\gamma_0$}
\end{center}
\par
\label{fig5}
\end{figure}

The influence of the dipole-dipole interaction is determined not only by the difference in absolute value of their resonance frequencies but also by the sign of the difference. For comparison in Fig. 5 we show two pairs of curves. The first one corresponds to the case when the resonance frequency of the embedded atom is greater than that of the surrounding atoms $\Delta=4\gamma_0$. The second group of curves is calculated for the same absolute value, but a negative detuning of $\Delta=-4\gamma_0$.  As the sign of the detuning changes, the sign of the line shift changes as well.

Such behavior of the level shift of an embedded atom can be illustrated by the example of diatomic system where the eigenstate problem is solved analytically \cite{44a}. For a system consisting of two different atoms the function $b_{0}(\omega)$ can be written as follows

\begin{equation} b_0(\omega)=\frac{1 }{f^{2}(r_{12})/(\omega+\Delta+i\gamma_{0}/2)-(\omega+i\gamma_{emb}/2)}
\label{7}
\end{equation}
In this formula $f(r_{12})$ is a function depending on the interatomic separation $r_{12}$.
\begin{equation} f(r_{12})=\frac{3}{4}\sqrt{\gamma_{0}\gamma_{emb}}\frac{1-ik_{0}r_{12}-
(k_{0}r_{12})^{2}}{(k_{0}r_{12})^{3}}\exp(ik_{0}r_{12}) \label{8}
\end{equation}

The solution (\ref{7})-(\ref{8}) is written in a coordinate frame with the Z-axis oriented along the interatomic axis and under the assumption that initially only one Zeeman sublevel $m=-1$ of the embedded atom is excited.

It can be seen from eqs. (\ref{7})-(\ref{8}) that for typical parameters of the considered systems the level shift decreases inversely with $\Delta$ for $|\Delta|>>\gamma_0/2$ and its sign changes  as the sign of $\Delta$ does.

\section{Conclusion}
In this paper, we analyze the influence of the dipole-dipole interatomic interaction on the process of spontaneous decay of atoms inside the cold atomic clouds under conditions when the averaged interatomic separation is less than, or comparable with, the wavelength of quasi resonance radiation. Besides decay dynamics, we analyze shifts of resonance as well as distortion of spectral shape of atomic transition. Two main cases are considered. First we study decay of atom identical to the surrounding ones. In addition we consider the case of an impurity atom, one with different properties than atoms of the surroundings.

The calculations were made on the basis of a quantum approach taking into account the vector nature of the electromagnetic field and Zeeman structure of atomic sublevels. A continuous medium approximation was not used. It allowed us to take into consideration random inhomogeneity of the atomic system and consequently the existing interatomic correlations.

It was shown that under the considered conditions, decay dynamics can be described by a multi exponential law. Instantaneous decay rate depends on time. At the beginning of the decay it is greater than decay constant of a free atom $\gamma$. In due course the rate decreases and becomes less than $\gamma$. We also found a noticeable distortion of the spectral shape of transition in comparison with the Lorentz profile typical for motionless free atoms.

Our calculation confirmed expected results consisting in increasing environmental influence as parameters of emitting atom approach those of surrounding atoms.  Most significantly this influence manifests itself when the initially excited atom is identical to the other ones. Particularly, in this case the distortion of the transition is most prominent and the shift of resonance is maximal. The latter becomes comparable with the half-width of atomic transition even for relatively small density $n=0.05$.

In our opinion results obtained in this paper are very important for future improvement in quantum frequency standards based on optical transitions in cold atomic ensembles. In such standards all atoms are identical to each other and for optimization of these devices the density dependence of the main characteristics have to be taken into account.

\section{Acknowledgements}
We thank M. D. Havey  for useful discussions. We acknowledge financial support from Ministry of Education and Science of Russian Federation (State Assignment № 3.1446.2014К)  and Russian Science Support Foundation. ASK appreciate the financial support from the Russian Foundation for Basic Research (Grant No. RFBR-14-02-31422), Russian President Grant for Young Candidates of Sciences (project MK-5318.2010.2), the Government of Saint-Petersburg, and the company "British Petroleum".

\end{document}